\pdfoutput=1
\documentclass[11pt,twoside,a4paper,cmspaper,final,collab]{cms-tdr}
\def\svnVersion{99f9425}\def\svnDate{2023/07/14}\def\cmsCernNoTag{CERN-EP-2022-190}\def\cmsCernDate{\today}\def\cmsMessage{Published in Physics Letters B as \href{http://dx.doi.org/10.1016/j.physletb.2023.137849}{\doi{10.1016/j.physletb.2023.137849}.}}
\begin{document}\cmsNoteHeader{HIN-20-003}

\hyphenation{back-ground-sub-traction}
\newcommand{\pttrk}{\ensuremath{\pt^{\text{trk}}}\xspace}
\newcommand{\Deltar}{\ensuremath{\Delta r}\xspace} 
\newcommand{\Rhodr}{\ensuremath{\rho(\Deltar)}\xspace}
\newlength\cmsTabSkip\setlength{\cmsTabSkip}{1ex}

\cmsNoteHeader{HIN-20-003}
\title{Search for medium effects using jets from bottom quarks in PbPb collisions at 
\texorpdfstring{$\sqrtsNN=5.02\TeV$}{sqrt(s[NN]) = 5.02 TeV}}

\date{\today}

\abstract{
The first study  of the shapes of jets arising from bottom ($\PQb$) quarks in heavy ion collisions is presented. Jet shapes are studied using charged hadron constituents as a function of their radial distance from the jet axis. Lead-lead (PbPb) collision data at a nucleon-nucleon center-of-mass energy of $\sqrtsNN=5.02\TeV$ were recorded by the CMS detector at the LHC, with an integrated luminosity of 1.69\nbinv. Compared to proton-proton collisions, a redistribution of the energy in \PQb jets to larger distances from the jet axis is observed in PbPb collisions. This medium-induced redistribution is found to be substantially larger for \PQb jets than for inclusive jets.
}

\hypersetup{
pdfauthor={CMS Collaboration},
pdftitle={Search for medium effects using jets from bottom quarks in PbPb collisions at 5.02 TeV},
pdfsubject={CMS},
pdfkeywords={CMS, beauty quark, b jets, heavy ion, QGP, jet quenching}}

\maketitle

\section{Introduction}

A hot dense partonic state of strongly interacting matter was discovered in relativistic heavy ion collisions over two decades ago~\cite{BRAHMS:2004adc,PHOBOS:2004zne,STAR:2005gfr,PHENIX:2004vcz}. Since then, its properties have been under intense study. One of the most exciting windows into this system (commonly called the quark-gluon plasma, QGP), and its properties is revealed by using jets as tomographic probes of the created medium. 
Jets, collimated streams of particles produced by hadronization of hard-scattered partons, have long been studied in proton-proton ($\Pp\Pp$) collisions. In the case of lead-lead (PbPb) collisions, energetic partons are known to lose energy while traversing the QGP, leading to modified jet spectra as compared to those from $\Pp\Pp$ collisions. 
Similarly, interactions of the initial hard-scattered and the fragmentation partons with the medium have been found to modify the final-state jet showers~\cite{ATLAS:2014ipv,CMS:2018zze,Bjorken:141477,dEnterria:2009xfs}. These phenomena, collectively called "jet quenching", provide invaluable input to theoretical calculations of medium properties. Substantial progress has been made in understanding the experimental results. Of particular interest  are unsettled questions involving details of the jet quenching, specifically the color-charge,  flavor and parton mass dependence of the energy loss. There are also open questions about the details of the medium response to the evolving jet. Thus, jets initiated  by bottom quarks (\PQb jets) provide unique experimental means to investigate the mass and parton flavor dependence of quenching effects. 
Recent measurements show that the suppression of heavy-flavor mesons and jets in heavy ion collisions, with respect to a $\Pp\Pp$ collision reference scaled by the average number of binary nucleon-nucleon interactions per PbPb collision, is different from that from inclusive (all flavors) measurements~\cite{ATLAS:2022fgb,ALICE:2021rxa,CMS:2018bwt}. 
These results confirm theoretical expectations that the energy loss depends on the type of parton involved. As jet shapes are especially sensitive to the details of the parton shower evolution, such measurements for \PQb jets provide an opportunity to  further explore parton transport and energy loss in the medium. 

The transverse momentum profile of charged particles in the jets is defined as
\begin{equation}\label{eq:P}
  P(\Deltar) = \frac{1}{\Deltar_{\mathrm{b}} - \Deltar_{\mathrm{a}}} \frac{1}{N_{\text{jet}}}\large{\Sigma_{\text{jets}}} \Sigma_{\text{trk} \in ( \Deltar_{\mathrm{a}}, \Deltar_{\mathrm{b}} )} \pttrk,
\end{equation}
where the radial distance between a track and the jet axis is defined in pseudorapidity ($\eta$) and azimuthal angle ($\phi$) as $\Deltar = \sqrt{(\eta^{\text{jet}}-\eta^{\text{trk}})^2+(\phi^{\text{jet}}-\phi^{\text{trk}})^2}$, $\Deltar_{\mathrm{a}}$ and $\Deltar_{\mathrm{b}}$ are the edges of rings in $\Deltar$, defining the radial binning for the analysis, and $\pttrk$ is the charged particle's  transverse momentum (\pt)  with respect to the beam. 
This profile is normalized to unity within the measured range of $\Deltar<1$ to produce the jet shape:
\begin{equation}\label{eq:jet_shape}
  \Rhodr = \frac{P(\Deltar)}{\large{\Sigma_{\text{jets}}} \Sigma_{\text{trk}\in (\Deltar<1)} \pttrk},
\end{equation}
which indicates how the \pt of charged particles is distributed with respect to the jet axis. Jet shapes are measured by the jet-track correlation technique, which can reliably separate the jet structure from the underlying event (UE) and extend the measurement from the jet cone region ($\Deltar<0.4$) to a larger $\Deltar$ region ($0.4<\Deltar <1)$~\cite{Aad:2014wha,CMS:2018zze}. 

This Letter reports the first bottom-quark jet shape measurement with charged particles studied in the QGP environment created by relativistic heavy ion collisions. The study is performed using PbPb collisions at a center-of-mass energy per nucleon-nucleon pair of $\sqrtsNN = 5.02\TeV$. Data, corresponding to an integrated luminosity of $1.69\nbinv$, were collected in 2018 using  the CMS detector~\cite{CMS-PAS-LUM-18-002, CMS:2021xjt, CMS:2008xjf} at the CERN LHC. Results reported in this Letter are also compared with shapes of inclusive jets, and with \PQb jet shapes measured in $\Pp\Pp$ collisions at the same center-of-mass energy~\cite{Sirunyan:2020pan}. Tabulated results are provided in the HEPData record for this analysis~\cite{HEPData}.

\section{CMS Detector}

The CMS apparatus is a multipurpose, nearly hermetic detector, designed to trigger on~\cite{CMS:2020cmk,CMS_run2_trigger_system} and identify electrons, muons, photons, and charged and neutral hadrons~\cite{CMS:2015xaf,CMS:2018rym,CMS:2015myp,CMS:2014pgm}. A global reconstruction "particle-flow" (PF) algorithm~\cite{CMS:2017yfk} combines the information provided by the all-silicon inner tracker and by the crystal electromagnetic, and brass and scintillator hadron calorimeters, operating inside a 3.8\unit{T} superconducting solenoid, with data from gas-ionization muon detectors interleaved with the solenoid flux-return yoke, to build \PGt leptons, jets, missing transverse momentum, and other physics objects~\cite{CMS:2018jrd,CMS:2016lmd,CMS:2019ctu}. 
Events of interest are selected using a two-tiered trigger system: a hardware-based level-one trigger, and High-Level Trigger, which uses partial online event processing for relevant physics objects (such as jets, in this case) ~\cite{CMS:2020cmk,CMS_run2_trigger_system}. 
The primary vertex (PV) is taken to be the vertex with the highest particle multiplicity, evaluated using tracking information alone, as described in Section 9.4 of Ref.~\cite{CMS-TDR-15-02}. The hadron forward  (HF) calorimeter uses steel as an absorber and quartz fibers as the sensitive material. The two halves of the HF are located 11.2\unit{m} from the interaction region, one on each side, and together they provide coverage in the range $3.0 < \abs{\eta} < 5.2$. The HF calorimeters are azimuthally subdivided into $20^\circ$ modular wedges and further segmented to form $0.175{\times}0.175$ ($\Delta \eta{\times}\Delta \phi$) ``towers".  They also serve as the luminosity monitors.

\section{Event Selection and Analysis Technique}

The PbPb data used in this analysis were selected with two calorimeter-based triggers that use an anti-\kt jet clustering algorithm~\cite{bib_antikt, Cacciari:2011ma} with a radius parameter of $R=0.4$. These two triggers require at least one jet with \pt greater than 80\GeV or 100\GeV, respectively,  after subtracting the UE background contribution iteratively~\cite{Kodolova:2007hd}. 
The data selected by these two triggers combined are used to extract the $P(\Deltar)$ signals. Additional data collected with a minimum bias trigger~\cite{CMS:2011aqh} are used to correct for the limited jet and track acceptance via an event mixing technique, which correlates jets with tracks from different events and thus produces a reference correlation containing only  detector and acceptance effects~\cite{CMS:2018zze}. Vertex and noise filters~\cite{Chatrchyan:2011sx, Chatrchyan:2012nia} are applied to reduce contamination of noncollision events, including calorimeter noise and beam-gas collisions. These filters require that at least one HF tower on each side of the interaction point have energy above 4\GeV  and that the PV position along the beam direction lies within 15\unit{cm} of the nominal interaction point. The collision centrality, expressed as a percentile of the total inelastic hadronic cross section, is determined based on the total energy deposited in both HF calorimeters~\cite{Chatrchyan:2012nia}. The events with 0\% centrality have the largest overlap of the two colliding nuclei.

Monte Carlo (MC) simulated event samples are used to evaluate the performance of the event reconstruction, the track reconstruction efficiency, and the jet energy response and resolution. The \PYTHIA (version 8.226) event generator~\cite{pythia82} with the CP5 tune~\cite{pythiaCP5tune} and parton distribution function set NNPDF3.1 at next-to-next-to-leading
order~\cite{nnpdf31} was used to simulate the hard scattering, the parton showering, and the hadronization of the partons, as well as the UE. For comparison to PbPb data, these PYTHIA events are embedded into samples from the \HYDJET event generator~\cite{hydjet} to simulate the background contribution for heavy ion events.  The jet flavors in simulations are determined at the event generator level by the presence of \PQb or $\PQc$ hadrons in the jet cone~\cite{CMS:2017wtu}. Those jets containing at least one \PQb hadron are defined as \PQb jets, therefore, the jets from the gluon splitting processes (GSP) which generate $\PQb\PAQb$ pairs are also considered as \PQb jets. The \GEANTfour~\cite{bib_geant} toolkit is used to simulate the CMS detector response. An additional reweighting procedure is performed to match the distribution of the PV along the beam direction to that observed in the data to ensure the best possible agreement between data and simulations.

In both data and simulation, charged particles are reconstructed using an iterative tracking method~\cite{CMS:2014pgm} based on hit information from  the silicon  subdetectors, permitting the reconstruction of charged particles within $\abs{\eta}<2.4$. The tracking efficiency ranges from approximately 60 to 90\%, depending on kinematic region and collision centrality, with  higher values in more peripheral events and for higher \pt particles. Tracks with $\pt>1\GeV$ and $\abs{\eta}<2.4$ are used in this study. 

Jets are reconstructed offline from PF candidates~\cite{CMS:2017yfk}, clustered using the anti-\kt algorithm with a radius parameter of $R=0.4$ and a constituents merging scheme that sums their 4-momentum directly (E-scheme).  
Simulation-derived corrections have been applied to the reconstructed jets to correct for the measured energy distortion arising from the limited detector resolution~\cite{CMS:2011shu,CMS:2016lmd}. Dedicated background subtraction iterations are applied to subtract the soft UE from the measured jet energy in PbPb collisions~\cite{CMS:2021nhn}. Furthermore, a different jet axis definition, known as the ``winner-takes-all" (WTA), is used for the jet shape analysis. The WTA axes are defined primarily by the leading constituents and reconstructed by reclustering the  E-scheme jets using the WTA combination rules~\cite{Bertolini:2013iqa,Larkoski:2014uqa}. 
The default E-scheme axis definition is not optimal for studying jet-track correlations because of an artificial feature that results in a substantial depletion of particles just outside the jet cone radius. The choice of the WTA axis definition avoids the need to correct for this artifact, thereby leading to smaller uncertainties.

A combined secondary vertex (CSV) discriminator is used to select \PQb jet candidates from the inclusive jets that have $\pt > 120\GeV$ and $\abs{\eta}<1.6$. This discriminator is a multivariate classifier that makes use of information of reconstructed secondary vertices  and the associated tracks to discriminate \PQb jets from charm and light-flavor quark jets, as well as gluon jets~\cite{Chatrchyan:2012jua, CMS:2014pgm,CMS:2017wtu}. The working point selected for this analysis leads to an overall 75\% \PQb jet selection efficiency and 45\% purity (i.e. true \PQb jet fraction of the tagged sample) as determined from an inclusive multijet sample. The selection is motivated by the balance between statistical and systematic uncertainties.

The \PQb jet shapes $\rho_{\text{b}}(\Deltar)$ are extracted following the jet-track correlation method~\cite{Sirunyan:2020pan} for six bins of \pttrk with boundaries 1, 2, 3, 4, 8, 12, and 300\GeV, which are the same as the binning scheme used in Ref~\cite{Sirunyan:2020pan}. Corrections for detector and jet-track pair acceptance effects are derived by using the event-mixing technique. The background is then subtracted from measured correlations in a data-driven manner using the measured signal in a data sideband far from the jet axis in a large-$\Delta\eta$ region~\cite{CMS:2018zze}. The purity of the \PQb jets is evaluated using MC simulations with adjustment by a data-to-MC difference evaluated from a negative-tagging technique~\cite{CMS:2017wtu}.
 The residual contamination of light-flavor and $\PQc$ jets mistagged as \PQb jets is removed using the templates derived from the measured inclusive jet shape signals $\rho_{\text{incl.}}(\Deltar)$, as described in Ref.~\cite{Sirunyan:2020pan}. 
Inefficiencies and nonlinearities in the tagging algorithm result in differences between the samples of all \PQb jets and tagged \PQb jets. The effects of this selection bias are estimated as functions of \pttrk, $\Deltar$, and centrality using the MC samples, and the resulting three-dimensional weighting is used to correct the raw \PQb jet shapes. 
Finally, simulation-based corrections are derived and applied following the method described in Refs.~\cite{CMS:2018zze,Khachatryan:2016tfj} to account for the jet axis resolution, tracking reconstruction efficiency, and background fluctuations.

\section{Systematic Uncertainties}

Several sources of systematic uncertainties are considered, including trigger efficiency, tracking efficiency, $\PQb$~tagging bias corrections, light flavor decontamination procedure, jet reconstruction, pair acceptance corrections, and underlying event background subtraction. The systematic uncertainties, summarized in Table~\ref{tab:sys} as a function of  centrality, are treated as uncorrelated, and the total systematic uncertainty is calculated by adding the contributions from individual sources in quadrature. The evaluation of each source of uncertainty is discussed below. 

\begin{table*}[htbp]
\centering
\topcaption{Systematic uncertainties in percentage for jet shape measurements for the various sources and collision centralities. Columns correspond to different centrality selections as marked. Where an uncertainty range is given, the upper edge of the range corresponds to the bin with the smallest \pttrk values. The sources from the decontamination and tagging bias are exclusive for \PQb jets. }

\begin{tabular}{lcccccc}
\hline
\multirow{3}{*}{Sources} & \multicolumn{3}{c}{$\PQb$ jets} & \multicolumn{3}{c}{Inclusive jets}\\
& \multicolumn{3}{c}{Centralities} & \multicolumn{3}{c}{Centralities}\\
& 30--90\% & 10-30\% & 0-10\% & 30--90\% & 10--30\% & 0--10\%\\
\hline
Trigger efficiency                       & 3.0         & 3.0         & 3.0         & 3.0      & 3.0       & 3.0\\
Tracking efficiency                      & 5.8         & 5.8         & 5.8         & 5.8      & 5.8       & 5.8\\
Tagging bias corrections                 & 5.0         & 5.0         & 5.0         & \NA      & \NA       & \NA\\
Decontamination procedure                & 8.0         & 8.0         & 8.0         & \NA      & \NA       & \NA\\
Jet energy scale/resolution                             & 4.2         & 4.2         & 4.2         & 4.2      & 4.2       & 4.2\\
Pair-acceptance corrections              & 1.0--2.0    & 1.0--4.0    & 1.0--5.0    & 1.0--2.0 & 1.0--3.0  & 1.0--4.0\\
Background subtraction                   & 1.0         & 2.0         & 3.0         & 1.0      & 2.0       & 3.0\\[\cmsTabSkip]
Total                                    & 12.3--12.4  & 12.3--12.9  & 12.3 --13.2 & 7.8--8.0 & 7.8--8.3  & 7.8--8.7\\
\hline
\end{tabular}
\label{tab:sys}
\end{table*}

The jet trigger is not fully efficient for jets with $\pt <160\GeV$ and  about 5\% of jets with $120 < \pt <160\GeV$ are lost in PbPb collisions. 
To estimate the uncertainty due to this trigger inefficiency, the analysis was repeated by using the data selected by a trigger with a threshold at 60\GeV. A 3\% uncertainty, which is the maximum observed difference relative to the nominal jet shape results, is assigned for this source. 

Comparing the tracks associated with inclusive jets, more tracks from $\PQb$~jets are far from a PV, which caused at most 4\% worse tracking reconstruction efficiency for $\PQb$~jets. The full magnitude of the observed difference is assigned as a conservative estimate of the tracking reconstruction uncertainty. An additional uncertainty of 4\% accounts for data-to-simulation track reconstruction differences, and is estimated from a study of \PD meson decays~\cite{CMS-PAS-TRK-10-002}. These two tracking-related uncertainties are added in quadrature, leading to a combined value of 5.8\%.

The fraction of $\PQb$~jets originating from the gluon splitting  mechanism in MC is less than what was observed in data by 15--20\%~\cite{CMS:2017wtu,Sirunyan:2018jju}. This is the dominant source of uncertainty for correcting the bias induced by the CSV discriminator. Varying the GSP fraction in the jet candidates by 20\% leads to less than a 5\% difference in the results, which is propagated as the uncertainty. Using the negative-tagging method, the uncertainty in the purity scale factor, caused by differences between data and simulated events, is estimated and ranges between 10--20\%, depending on the jet \pt region. This uncertainty is propagated to the final observables, resulting in an 8\% uncertainty for the decontamination procedure described in Ref.~\cite{Sirunyan:2020pan}. The overall jet energy scale (JES) is sensitive to the sample's relative fraction of quark and gluon jets. The differences of the JES for quark and gluon jets were studied, and the jet energy scaling for inclusive jets has been varied by the largest observed differences in the JES corrections.  For b-tagged jets, the largest observed difference between inclusive and tagged jets was used.
The differences found in the measured correlation distributions, defined by Eq.~\ref{eq:jet_shape},  are assigned as the systematic uncertainties due to this quark/gluon effect. These uncertainties are found to be below 3\%, since the in-jet multiplicity and the jet fragmentation function change slowly with the jet \pt. A similar procedure is used for the b-to-inclusive jet shape ratios, and the observed 1.5\% difference is assigned as the JES uncertainty. The jet energy resolution (JER) data-to-simulation difference, based on $\gamma$+jet studies~\cite{CMS:2011shu}, is 20\% or below. The corresponding uncertainty was evaluated by smearing the reconstructed jet \pt to worsen the JER by the maximum value of 20\% and repeating the study. The resulting yield variation compared to the nominal results was found to be below 3\%. A similar study is made for b-to-inclusive jet shape ratios. It shows that the JER uncertainty is fully correlated and hence is canceled in the ratios. In total, the JES- and JER-related systematic uncertainties assigned for the \PQb jet shapes (extracted using the particle yields) and b-to-inclusive ratios are 4.2 and 1.5\%, respectively. 

The uncertainties of the pair acceptance correction are estimated from the difference between positive and negative $\Delta\eta$ sideband regions. Theoretically, there should be no $\Delta\eta$ dependence in the regions $1.5<\abs{\Delta\eta}<2.5$, which are far away from the jet axis. Any deviations from this expectation are used to quantify the related systematic error. This uncertainty amounts to 1.0--5.0\%, depending on the centrality, the track $\pttrk$ bin, and the $\Deltar$ bin. Uncertainties associated with the background subtraction are evaluated by considering the variations between the two sideband regions ($1.5 < \abs{\Delta\eta} < 2.0$ and $2.0 < \abs{\Delta\eta} < 2.5$) on each side of the detector after the background subtraction. The statistical uncertainties on the background level determination are also added in quadrature to get the final background subtraction uncertainty ranging from 1.0 to 3.0\%, depending on the centrality.

\section{Results}

\begin{figure*}[p!]
  \centering
     \includegraphics[width=0.86\textwidth]{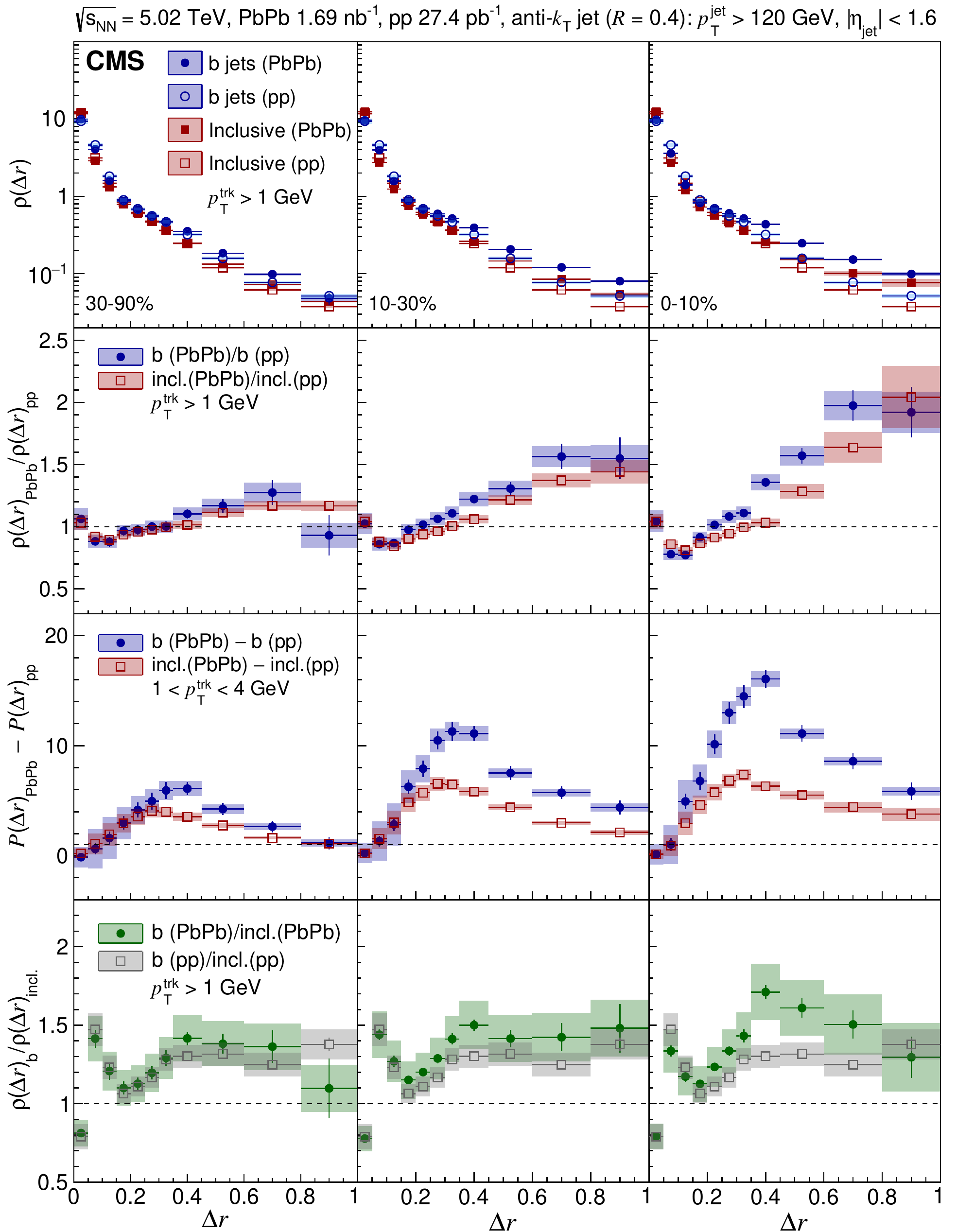}
       \caption{Upper row: jet shape distributions $\rho(\Deltar)$ for inclusive (solid red squares) and \PQb jets (solid blue circles) with $\pt >120\GeV$ in three centrality bins (left to right) for PbPb collisions at $\sqrtsNN=5.02\TeV$, as well as in $\Pp\Pp$ collisions (open red and blue symbols) from Ref.~\cite{Sirunyan:2020pan}, which are identical in all three panels. Second row: ratio of PbPb to $\Pp\Pp$ jet shape results for inclusive (red) and \PQb jets (blue). Third row: the difference between the charged-particle transverse momentum profile, defined by Eq.~\ref{eq:P}, between PbPb and $\Pp\Pp$ collisions  for $1<\pttrk<4\GeV$ for inclusive (red) and \PQb jets (blue). Lower row: ratio of \PQb to inclusive jet shapes for several PbPb centrality bins (green), as well as $\Pp\Pp$ collisions from Ref.~\cite{Sirunyan:2020pan} (grey), which are identical in all three panels. In all panels, the vertical bars and shaded boxes represent the statistical and systematic uncertainties, respectively.}
     \label{fig:results}
\end{figure*}

The shapes of \PQb jets and inclusive jets, measured using charged particles with $\pttrk > 1\GeV$ from PbPb collisions, are shown in the upper row of Fig.~\ref{fig:results}. For comparison, the results from $\Pp\Pp$ collisions in Ref.~\cite{Sirunyan:2020pan} are also shown in each panel. The \PQb jets are observed to have a broader shape than that of inclusive jets.  In \PYTHIA studies, this difference could be primarily explained by contributions from the GSP  producing \PQb jet pairs. The ratios of jet shapes measured in PbPb collisions to those from $\Pp\Pp$ data from Ref.~\cite{Sirunyan:2020pan} are shown in the second row of Fig.~\ref{fig:results}. Both the \PQb and inclusive jet shape ratios show a redistribution of the transverse momentum of jet constituents from small to large distances from the jet axis. This is evident in the depletion of charged-particle tracks in the range of about $0.05 \lesssim \Deltar \lesssim 0.4$, followed by a strong enhancement above $\Deltar \approx 0.4$ seen for both ratios. For PbPb collisions, the observed large $\Deltar$ enhancements are centrality dependent and are most significant in central collisions, consistent with expectations based on jet-medium interactions for partons traversing the QGP. Figure~\ref{fig:results} shows an even larger PbPb to $\Pp\Pp$ ratio for \PQb jets than for inclusive jets in the large-$\Deltar$ region. 
This QGP-induced excess of transverse momentum at large $\Deltar$ could result from the response of the medium to the propagating jet,  with the excited portion of the medium  hadronizing into additional particles around the jet direction. The presented measurement may then imply that \PQb jets can cause a larger medium response than inclusive jets.
We find that charged particles with $1 <\pttrk < 4\GeV$ carry almost all of the observed momentum excess. The differences of the transverse momentum profiles (as opposed to the ratios of the shapes) between the PbPb and $\Pp\Pp$ measurements characterize the magnitude of the measured excess momentum and are plotted in the third row of Fig.~\ref{fig:results} for both the \PQb and inclusive jets. Unlike b-to-inclusive jet shape ratios, which could be affected by the changes in both the numerator and denominator, the difference of the momentum profiles represents an absolute change in the momentum from $\Pp\Pp$ to PbPb data. A more significant absolute momentum excess in PbPb collisions with respect to the $\Pp\Pp$ reference is found for \PQb jets than for inclusive jets. 

To quantify the difference between the shapes of \PQb and inclusive jets within a given system, the lower row of Fig.~\ref{fig:results} presents the shape ratios for \PQb jets divided by inclusive jets for both $\Pp\Pp$ and different centralities of PbPb collisions. Previous studies in $\Pp\Pp$ collisions~\cite{Sirunyan:2020pan} have demonstrated that the \PQb jet shapes are broader than inclusive jet shapes in the case of vacuum fragmentation (no QGP). For comparison, PbPb ratios are overlaid with the $\Pp\Pp$ baseline measurement from Ref.~\cite{Sirunyan:2020pan}. These ratios also show a centrality-dependent enhancement of the large-$\Deltar$ distributions for \PQb over inclusive jets. The small $\Deltar$ region shows another interesting finding: a depletion is seen in the \PQb to inclusive jet ratios for roughly $\Deltar \lesssim 0.05$  for all centrality bins without apparent centrality dependence. It is argued that in $\Pp\Pp$ collisions this dip could be interpreted as a manifestation of the "dead-cone" effect, resulting from a larger suppression of collinear parton radiation from a heavy bottom quark than from light partons~\cite{Sirunyan:2020pan} . Comparing PbPb to $\Pp\Pp$ results, no significant difference is seen in this small $\Deltar$ dead-cone-like structure at any centrality.

\section{Summary}

In summary, the first jet shape measurement for bottom quark ($\PQb$) jets in lead-lead (PbPb) collisions at a nucleon-nucleon center-of-mass energy of $\sqrtsNN=5.02\TeV$ are presented. 
A jet-track correlation technique was used with \PQb jets of $\pt > 120\GeV$ and charged particles of transverse momenta $\pttrk>1\GeV$ to measure jet shapes as a function of radial distance $\Deltar$ between the jet axes and particle tracks.
The measured \PQb jet shapes show a depletion of transverse momenta at small radial distances from the jet axis compared to inclusive jet shapes. This depletion is already present in proton-proton ($\Pp\Pp$) events at the same center-of-mass energy and does not notably change looking at PbPb collisions at different centralities. This observation may provide a quantitative measurement of the expected dead-cone effect for \PQb jets. Comparisons of jet shapes from PbPb and $\Pp\Pp$ data show that the presence of the quark-gluon plasma modifies the energy flow around \PQb jets. These modifications include a depletion of the transverse momenta in the range of about $\Deltar \lesssim 0.4$ and an enhancement at larger radial distances from the jet axis. Furthermore, the large $\Deltar$ enhancement is found to be 
greater than that previously reported for inclusive jets in PbPb collisions. This observation is consistent with an increased medium response to the propagation of a heavier quark. The enhancement is centrality dependent and is largest in the most central collisions. This measurement provides new constraints for theoretical calculations of parton flavor dependence of energy loss and jet-medium interactions in the quark-gluon plasma.

\begin{acknowledgments}
We congratulate our colleagues in the CERN accelerator departments for the excellent performance of the LHC and thank the technical and administrative staffs at CERN and at other CMS institutes for their contributions to the success of the CMS effort. In addition, we gratefully acknowledge the computing centers and personnel of the Worldwide LHC Computing Grid and other centers for delivering so effectively the computing infrastructure essential to our analyses. Finally, we acknowledge the enduring support for the construction and operation of the LHC, the CMS detector, and the supporting computing infrastructure provided by the following funding agencies: BMBWF and FWF (Austria); FNRS and FWO (Belgium); CNPq, CAPES, FAPERJ, FAPERGS, and FAPESP (Brazil); MES and BNSF (Bulgaria); CERN; CAS, MoST, and NSFC (China); MINCIENCIAS (Colombia); MSES and CSF (Croatia); RIF (Cyprus); SENESCYT (Ecuador); MoER, ERC PUT and ERDF (Estonia); Academy of Finland, MEC, and HIP (Finland); CEA and CNRS/IN2P3 (France); BMBF, DFG, and HGF (Germany); GSRI (Greece); NKFIH (Hungary); DAE and DST (India); IPM (Iran); SFI (Ireland); INFN (Italy); MSIP and NRF (Republic of Korea); MES (Latvia); LAS (Lithuania); MOE and UM (Malaysia); BUAP, CINVESTAV, CONACYT, LNS, SEP, and UASLP-FAI (Mexico); MOS (Montenegro); MBIE (New Zealand); PAEC (Pakistan); MES and NSC (Poland); FCT (Portugal); MESTD (Serbia); MCIN/AEI and PCTI (Spain); MOSTR (Sri Lanka); Swiss Funding Agencies (Switzerland); MST (Taipei); MHESI and NSTDA (Thailand); TUBITAK and TENMAK (Turkey); NASU (Ukraine); STFC (United Kingdom); DOE and NSF (USA).
    
\hyphenation{Rachada-pisek} Individuals have received support from the Marie-Curie program and the European Research Council and Horizon 2020 Grant, contract Nos.\ 675440, 724704, 752730, 758316, 765710, 824093, 884104, and COST Action CA16108 (European Union); the Leventis Foundation; the Alfred P.\ Sloan Foundation; the Alexander von Humboldt Foundation; the Belgian Federal Science Policy Office; the Fonds pour la Formation \`a la Recherche dans l'Industrie et dans l'Agriculture (FRIA-Belgium); the Agentschap voor Innovatie door Wetenschap en Technologie (IWT-Belgium); the F.R.S.-FNRS and FWO (Belgium) under the ``Excellence of Science -- EOS" -- be.h project n.\ 30820817; the Beijing Municipal Science \& Technology Commission, No. Z191100007219010; the Ministry of Education, Youth and Sports (MEYS) of the Czech Republic; the Hellenic Foundation for Research and Innovation (HFRI), Project Number 2288 (Greece); the Deutsche Forschungsgemeinschaft (DFG), under Germany's Excellence Strategy -- EXC 2121 ``Quantum Universe" -- 390833306, and under project number 400140256 - GRK2497; the Hungarian Academy of Sciences, the New National Excellence Program - \'UNKP, the NKFIH research grants K 124845, K 124850, K 128713, K 128786, K 129058, K 131991, K 133046, K 138136, K 143460, K 143477, 2020-2.2.1-ED-2021-00181, and TKP2021-NKTA-64 (Hungary); the Council of Science and Industrial Research, India; the Latvian Council of Science; the Ministry of Education and Science, project no. 2022/WK/14, and the National Science Center, contracts Opus 2021/41/B/ST2/01369 and 2021/43/B/ST2/01552 (Poland); the Funda\c{c}\~ao para a Ci\^encia e a Tecnologia, grant CEECIND/01334/2018 (Portugal); the National Priorities Research Program by Qatar National Research Fund; MCIN/AEI/10.13039/501100011033, ERDF ``a way of making Europe", and the Programa Estatal de Fomento de la Investigaci{\'o}n Cient{\'i}fica y T{\'e}cnica de Excelencia Mar\'{\i}a de Maeztu, grant MDM-2017-0765 and Programa Severo Ochoa del Principado de Asturias (Spain); the Chulalongkorn Academic into Its 2nd Century Project Advancement Project, and the National Science, Research and Innovation Fund via the Program Management Unit for Human Resources \& Institutional Development, Research and Innovation, grant B05F650021 (Thailand); the Kavli Foundation; the Nvidia Corporation; the SuperMicro Corporation; the Welch Foundation, contract C-1845; and the Weston Havens Foundation (USA).
\end{acknowledgments}

\bibliography{auto_generated} 

\cleardoublepage \appendix\section{The CMS Collaboration \label{app:collab}}\begin{sloppypar}\hyphenpenalty=5000\widowpenalty=500\clubpenalty=5000
\cmsinstitute{Yerevan Physics Institute, Yerevan, Armenia}
{\tolerance=6000
A.~Tumasyan\cmsAuthorMark{1}\cmsorcid{0009-0000-0684-6742}
\par}
\cmsinstitute{Institut f\"{u}r Hochenergiephysik, Vienna, Austria}
{\tolerance=6000
W.~Adam\cmsorcid{0000-0001-9099-4341}, J.W.~Andrejkovic, T.~Bergauer\cmsorcid{0000-0002-5786-0293}, S.~Chatterjee\cmsorcid{0000-0003-2660-0349}, K.~Damanakis\cmsorcid{0000-0001-5389-2872}, M.~Dragicevic\cmsorcid{0000-0003-1967-6783}, A.~Escalante~Del~Valle\cmsorcid{0000-0002-9702-6359}, P.S.~Hussain\cmsorcid{0000-0002-4825-5278}, M.~Jeitler\cmsAuthorMark{2}\cmsorcid{0000-0002-5141-9560}, N.~Krammer\cmsorcid{0000-0002-0548-0985}, L.~Lechner\cmsorcid{0000-0002-3065-1141}, D.~Liko\cmsorcid{0000-0002-3380-473X}, I.~Mikulec\cmsorcid{0000-0003-0385-2746}, P.~Paulitsch, F.M.~Pitters, J.~Schieck\cmsAuthorMark{2}\cmsorcid{0000-0002-1058-8093}, R.~Sch\"{o}fbeck\cmsorcid{0000-0002-2332-8784}, D.~Schwarz\cmsorcid{0000-0002-3821-7331}, M.~Sonawane\cmsorcid{0000-0003-0510-7010}, S.~Templ\cmsorcid{0000-0003-3137-5692}, W.~Waltenberger\cmsorcid{0000-0002-6215-7228}, C.-E.~Wulz\cmsAuthorMark{2}\cmsorcid{0000-0001-9226-5812}
% ...
\par}
\vskip\cmsinstskip
\dag:~Deceased\\
$^{1}$Also at Yerevan State University, Yerevan, Armenia\\
$^{2}$Also at TU Wien, Vienna, Austria\\
% ...
\end{sloppypar}
%%% END EDITABLE REGION %%%
% skeleton_end
\end{document}